\documentclass[prl,aps,twocolumn,floatfix,showpacs]{revtex4}
\usepackage{graphics}
\begin{document}
\title{Evolution from BCS to BEC superfluidity in $p$-wave Fermi gases}
\author{M. Iskin and C. A. R. S{\'a} de Melo}
\affiliation{School of Physics, Georgia Institute of Technology, Atlanta, Georgia 30332, USA}
\date{\today}

\begin{abstract}
We consider the evolution of superfluid properties of a three dimensional $p$-wave Fermi gas
from weak (BCS) to strong (BEC) coupling as a function of scattering volume.
We analyse the order parameter, quasi-particle excitation spectrum, 
chemical potential, average Cooper pair size and the momentum distribution
in the ground state ($T = 0$). 
We also discuss the critical temperature $T_{\rm c}$, chemical potential and number of 
unbound, scattering and bound fermions in the normal state ($T = T_{\rm c}$).
Lastly, we derive the time-dependent Ginzburg-Landau equation for $T \approx T_{\rm c}$
and extract the Ginzburg-Landau coherence length.

\end{abstract}
\maketitle

Arguably the next frontier of research in ultracold Fermi systems is the search 
for superfluidity in higher angular momentum states.
Substantial experimental progress has been made recently~\cite{regal,ticknor,zhang,schunck,gunter} 
in connection to $p$-wave cold Fermi gases, making them ideal candidates 
for the observation of novel triplet superfluid phases.
These phases may be present not only in atomic, 
but also in nuclear (pairing in nuclei), astrophysics (neutron stars), 
and condensed matter (organic superconductors) systems.

The tuning of $p$-wave interactions in ultracold Fermi gases was initially explored
via $p$-wave Feshbach resonances in trap geometries
for $^{40} {\rm K}$~\cite{regal,ticknor} and $^6{\rm Li}$~\cite{zhang,schunck}.
Finding and sweeping through these resonances is difficult
since they are much narrower than the $s$-wave ($\ell = 0$) case,
because atoms interacting via higher angular momentum channels ($\ell \ne 0$) have to
tunnel through a centrifugal barrier to couple to the bound state~\cite{ticknor}.
Furthermore, while losses due to two-body dipolar~{\cite{zhang,john}}
or three-body~\cite{regal,ticknor} processes
challenged earlier $p$-wave experiments, these losses were still present but were less dramatic
in the very recent optical lattice experiment involving $^{40} {\rm K}$
and $p$-wave Feshbach resonances~\cite{gunter}.

For a dilute $^{40}{\rm K}$ Fermi gas, the magnetic dipole-dipole interactions 
between valence electrons split $p$-wave ($\ell = 1$) Feshbach resonances that belong 
to different $m_\ell$ states~\cite{ticknor}. Therefore, the ground state is highly dependent on
the detuning and separation of these resonances, and 
possible $p$-wave superfluid phases can be studied from the Bardeen-Cooper-Schrieffer (BCS)
to the Bose-Einstein condensation (BEC) regime. For
instance, it has been proposed~\cite{gurarie,skyip} for sufficiently large splittings 
that pairing occurs only in $m_\ell = 0$ and does not occur in $m_\ell = \pm 1$ state,
while for small splittings, pairing occurs via a linear 
combination of the $m_\ell = 0$ and $m_\ell = \pm 1$ states.
Thus, these resonances may be tuned and studied independently if the
splitting is large enough in comparison to the experimental resolution.

The BCS to BEC evolution in $p$-wave systems was recently discussed at $T = 0$ 
for a two-hyperfine state (THS)~\cite{tlho} in three dimensions (3D),
and for a single-hyperfine state (SHS)~\cite{botelho,iskin} in two dimensions, using fermion-only models.
Furthermore, fermion-boson models were proposed to describe $p$-wave superfluidity
at zero~\cite{gurarie,skyip} and finite temperature~\cite{ohashi} in three dimensions.
Unlike the previous models, we present a zero and finite temperature analysis of 
SHS $p$-wave Fermi gas in 3D within a fermion-only description, where molecules 
naturally appear as bound states of two-fermions.

The Hamiltonian for a dilute SHS $p$-wave Fermi gas in 3D is given by
\begin{eqnarray}
\label{eqn:hamiltonian.pwave}
H=\sum_{\mathbf{k}}\xi(\mathbf{k})a_{\mathbf{k},\uparrow}^\dagger a_{\mathbf{k},\uparrow} + 
\frac{1}{2}\sum_{\mathbf{k},\mathbf{k'},\mathbf{q}}V_{\rm p} (\mathbf{k},\mathbf{k'})
b_{\mathbf{k},\mathbf{q}}^\dagger b_{\mathbf{k'},\mathbf{q}}, 
\end{eqnarray}
where the pseudo-spin $\uparrow$ labels the hyperfine state represented by 
the creation operator $ a_{\mathbf{k},\uparrow}^\dagger$, and
$b_{\mathbf{k},\mathbf{q}}^\dagger=a_{\mathbf{k}+\mathbf{q}/2,\uparrow}^\dagger 
a_{-\mathbf{k}+\mathbf{q}/2,\uparrow}^\dagger$.
Here, $\xi(\mathbf{k})= \epsilon(\mathbf{k}) - \mu$, where 
$\epsilon(\mathbf{k}) = k^2/(2M)$ is the energy of the fermions
and $\mu$ is the chemical potential.
The attractive interaction can be written in a separable form as
$
V_p(\mathbf{k},\mathbf{k'})= -4\pi g \Gamma^*(\mathbf{k})\Gamma(\mathbf{k'}) 
$
where $g > 0$. The function
$
\Gamma(\mathbf{k})= \Gamma_{\rm k} (k) \Gamma_{\rm a} (\hat{\mathbf{k}})
$
is a symmetry factor where $\Gamma_{\rm k}(k) = (k k_0) / (k^2 + k_0^2)$, and
$\Gamma_{\rm a} (\hat{\mathbf{k}}) = Y_{1,0}(\hat{\mathbf{k}})$ is the angular dependence.
In addition, $k_0 \sim R_0^{-1}$, where $R_0$ the interaction range in real space, sets the momentum scale.
Furthermore, the diluteness condition ($n R_0^3 \ll 1$)
requires $(k_0/k_{\rm F})^3 \gg 1$, where $n$ is the density of atoms
and $k_{\rm F}$ is the Fermi momentum.

In the imaginary-time functional integral formalism ($\hbar = k_{\rm B} = 1$ and $\beta=1/T$), 
the partition function can be written as 
$
Z=\int D[a^\dagger,a]e^{-S}
$
with an action given by
$
S=\int_0^\beta d\tau\left[ \sum_{\mathbf{k}}a_{\mathbf{k},\uparrow}^\dagger(\tau)(\partial_\tau) a_{\mathbf{k},\uparrow}(\tau) 
+ H(\tau) \right].
$
We first introduce the Nambu spinor 
$\psi^\dagger(p)=( a_{{\rm p}, \uparrow}^\dagger , a_{-{\rm p}, \uparrow} )$, 
and define $p=(\mathbf{k},w_\ell)$ 
to denote both momentum and fermionic Matsubara frequency $w_\ell=(2\ell+1)\pi/\beta$. 
Furthermore, we use the  Hubbard-Stratonovich transformation to decouple fermionic and bosonic degrees of freedoms.
Then, we integrate over the fermionic part, and rewrite the bosonic field
as a combination of $\tau$-independent $\Delta_0$ and $\tau$-dependent $\Lambda(q)$.
Here, $q=(\mathbf{q},v_\ell)$ with bosonic Matsubara frequency $v_\ell=2\ell\pi/\beta$.

Performing an expansion of $S$ to quadratic order in $\Lambda(q)$, we obtain
\begin{equation}
S_{\rm gauss} = S_0 + \frac{\beta}{2}\sum_{q} \bar{\Lambda}^\dagger(q) \mathbf{F}^{-1}(q) \bar{\Lambda}(q),
\end{equation}
where the vector $\bar{\Lambda}^\dagger(q)$ is such that
$\bar{\Lambda}^\dagger(q) = [\Lambda^\dagger(q), \Lambda(-q)]$,
and the matrix $\mathbf{F}^{-1}(q)$ is the inverse fluctuation propagator. 
Here, $S_0$ is the saddle point action given by
$
S_0 = \beta|\Delta_0|^2/(8\pi g) + 
\sum_{p}\left[\beta\xi(\mathbf{k})/2 - {\rm Tr}\ln(\beta\mathbf{G_0}^{-1}/2)\right], 
$
where the inverse Nambu propagator is
$
\mathbf{G}_0^{-1}=iw_\ell \sigma_0-\xi(\mathbf{k})\sigma_3 + 
\Delta_0^* \Gamma(\mathbf{k})\sigma_- + \Gamma^*(\mathbf{k}) \Delta_0 \sigma_+.
$
The fluctuation term in the action leads to a correction
to the thermodynamic potential, which can be written as 
$\Omega_{{\rm gauss}} = \Omega_0 + \Omega_{{\rm fluct}}$ with 
$\Omega_0 = S_0/\beta$ and
$\Omega_{{\rm fluct}} = \beta^{-1}\sum_{q}\ln\det[\mathbf{F}^{-1}(q)/(2\beta)]$.

The saddle point condition $\delta S_0 /\delta \Delta_0^* = 0$ leads to an equation for the order parameter
\begin{equation}
\frac{1}{4\pi g}=\sum_{\mathbf{k}}\frac{|\Gamma(\mathbf{k})|^2} 
{2E(\mathbf{k})} \tanh\frac{\beta E(\mathbf{k})}{2},
\end{equation}
where
$
E(\mathbf{k})=(\xi^2(\mathbf{k})+|\Delta(\mathbf{k})|^2)^{\frac{1}{2}}
$
is the quasi-particle energy, 
and
$
\Delta(\mathbf{k})= \Delta_0\Gamma(\mathbf{k})
$
is the order parameter.
The scattering amplitude within a T-matrix formulation~\cite{tlho} is
$
f (k) = k^2 / (- 1/a_{\rm p} + r_{\rm p} k^2 - ik^3) 
$
for the $p$-wave channel, where $a_{\rm p}$ is the scattering volume, and $r_{\rm p}$ has
dimensions of inverse length.
Using $f (k)$, we can elliminate $g$ in favor of $a_{\rm p}$ via the relation
\begin{equation}
\frac{1}{4\pi g} = -\frac{M V}{16\pi^2 a_{\rm p}k_0^2} 
+ \sum_{\mathbf{k}} \frac{|\Gamma(\mathbf{k})|^2}{2\epsilon(\mathbf{k})},
\end{equation}
where $V$ is the volume.

The order parameter equation has to be solved self-consistently with the number equation 
$N = -\partial \Omega/\partial {\mu}$ which leads to two contributions 
to the number equation $N = N_0 + N_{{\rm fluct}}$. 
$N_0 = -\partial \Omega_0/\partial {\mu}$ is the saddle point number equation given by
\begin{equation}
N_0 = \sum_{\mathbf{k}} n_0(\mathbf{k}); \, \, \, 
n_0(\mathbf{k})=\frac{1}{2} - \frac{\xi(\mathbf{k})}{2E(\mathbf{k})} \tanh\frac{\beta E(\mathbf{k})}{2},
\label{eqn:numbereqn}
\end{equation}
where $n_0(\mathbf{k})$ is the momentum distribution.
Similarly, $N_{{\rm fluct}} = -\partial \Omega_{{\rm fluct}}/\partial \mu$ is the 
fluctuation number equation given by
$
N_{{\rm fluct}} = -\beta^{-1}\sum_{q} \lbrace \partial [\det \mathbf{F}^{-1}(q)] / \partial {\mu} 
\rbrace / \det \mathbf{F}^{-1}(q).
$

\begin{figure} [htb]
\centerline{\scalebox{0.37}{\includegraphics{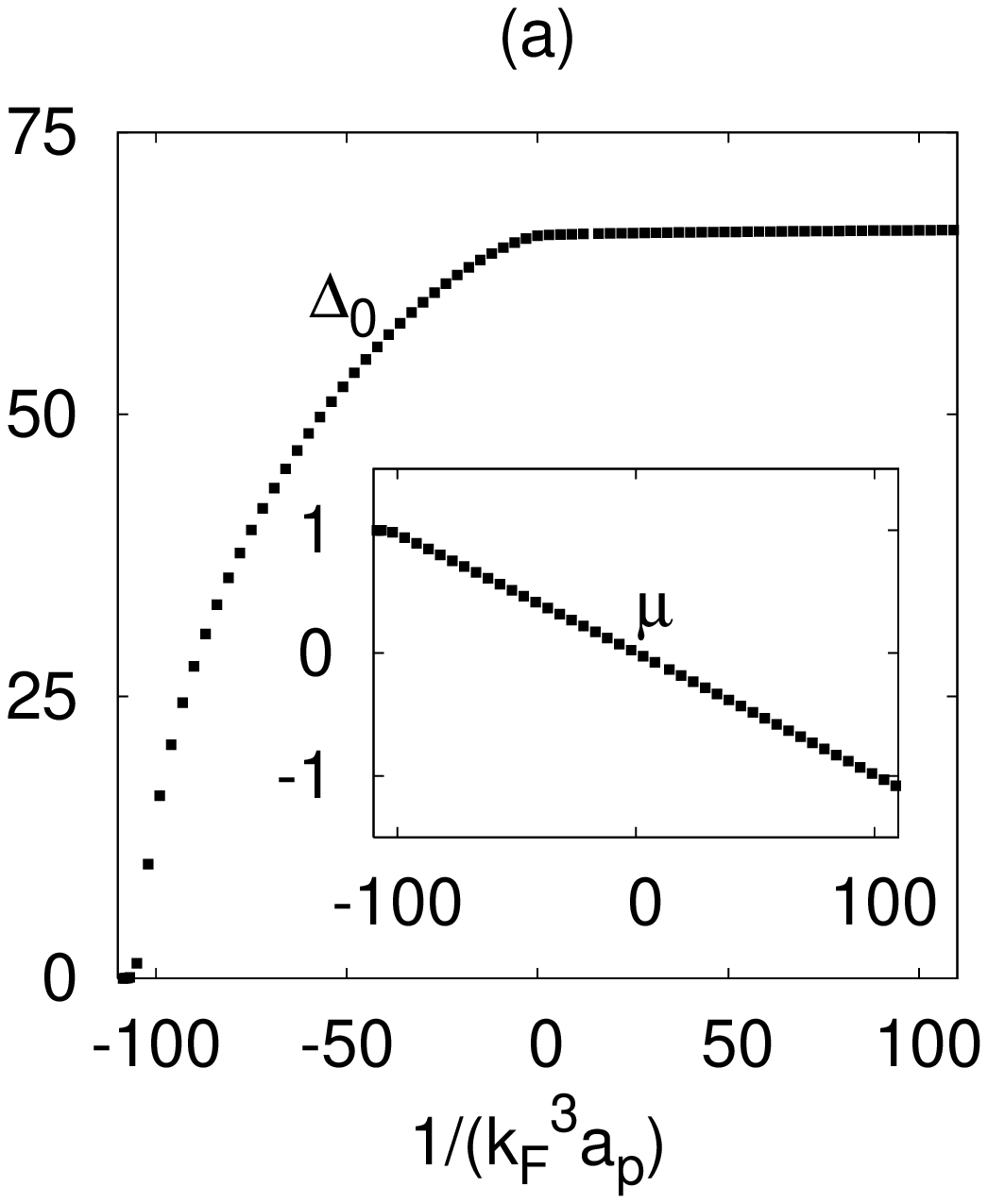} \includegraphics{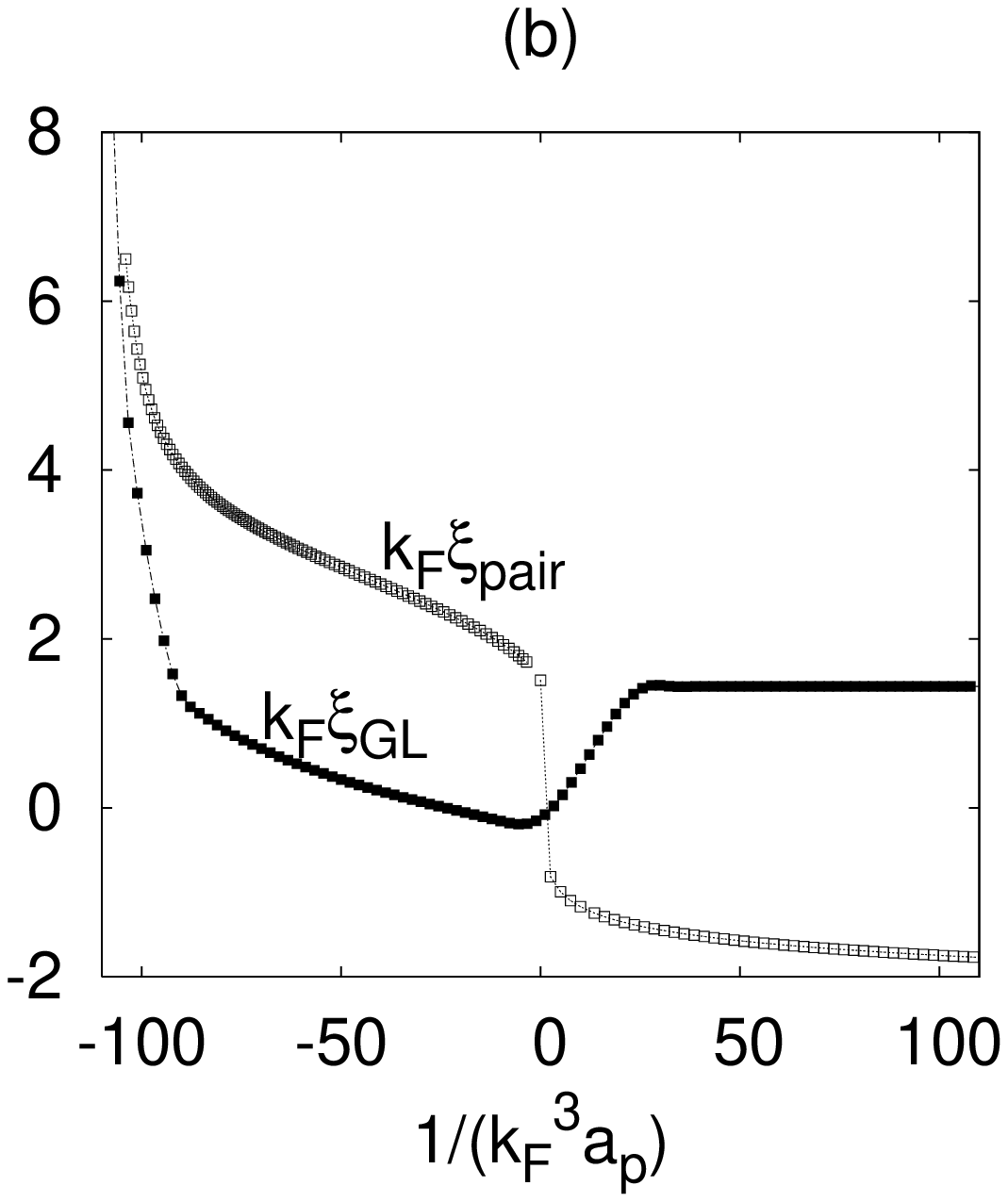}}}
\caption{\label{fig:gap.and.mu} Plots of reduced
a) order parameter amplitude $\Delta_{\rm r} = \Delta_0/\epsilon_{\rm F}$
and chemical potential $\mu_r = \mu/\epsilon_F$, and
b) average Cooper pair size $k_{\rm F}\xi_{{\rm pair}}$ at $T = 0$ and 
GL coherence length $k_{\rm F}\xi_{{\rm GL}}^{zz}$ at $T = T_{\rm c}$
in a logarithmic scale versus $1/(k_{\rm F}^3 a_{\rm p})$.
}
\end{figure}
For $T \approx 0$, $N_{\rm fluct}$ is small ($\propto T^4$)
compared to $N_0$~\cite{carlos} for any interaction strength leading to $N \approx N_0$.
In  Fig.~\ref{fig:gap.and.mu}a, we plot $\Delta_{\rm r} = \Delta_0/\epsilon_{\rm F}$ and 
$\mu_{\rm r} = \mu/\epsilon_{\rm F}$ at $T = 0$  as a function of 
$1/(k_{\rm F}^3 a_{\rm p})$,
where $\epsilon_{\rm F} = k_{\rm F}^2/(2M)$ is the Fermi energy.
Here, we choose $k_0 \approx 200 k_{\rm F}$. 
Notice that the BCS to BEC evolution range in $1/(k_{\rm F}^3 a_{\rm p})$ 
is $\sim k_0/k_{\rm F}$.
The weak coupling $\mu = \epsilon_{\rm F}$ changes continuously to the 
strong coupling $\mu = -1/(M k_0 a_{\rm p})$ when $k_0^3 a_{\rm p} \gg 1$. 
In strong coupling, $a_{\rm p}$ has to be larger than $a_{\rm p} > 2/k_0^3$ 
for the order parameter equation to have a solution
with $\mu < 0$, which reflects the Pauli exclusion principle.
In addition, the weak coupling
$\Delta_{0} = 24 (k_0/k_{\rm F})\epsilon_{\rm F} \exp[- 8/3 + \pi k_0 / (4k_{\rm F}) - \pi/(2k_{\rm F}^3 |a_p|)]$ 
evolves continuously to a constant
$\Delta_{0} = 8\epsilon_{\rm F} [\epsilon_0/(9\epsilon_{\rm F})]^{1/4}$
in strong coupling, where $\epsilon_0 = k_0^2/(2M)$.
The evolution of $\Delta_0$ and $\mu$ are qualitatively 
similar to recent $T = 0$ results for THS fermion~\cite{tlho} and SHS fermion-boson~\cite{skyip} models.
Due to the angular dependence of $\Delta(\mathbf{k})$,
the quasi-particle spectrum $E(\mathbf{k})$ is gapless ($\min E(\mathbf{k}) = 0$) for $\mu > 0$, 
and fully gapped $(\min E(\mathbf{k}) = |\mu|)$ for $\mu < 0$. 
Furthermore, both $\Delta_0$ and $\mu$ are non-analytic exactly when $\mu$
crosses the bottom of the fermion energy band $\mu = 0$ at $1/(k_{\rm F}^3 a_p) \approx 0.5$.
The non-analyticity does not occur in the first derivative of $\Delta_0$ or $\mu$
as it is the case in 2D~\cite{botelho}, but occurs in the second and higher derivatives.
Therefore, the evolution from BCS to BEC is not smooth, 
and a topological gapless to gapped quantum phase transition~\cite{botelho,gurarie} 
takes place when $\mu = 0$.

\vskip 10mm
\begin{figure} [htb]
\centerline{\scalebox{0.37}{\includegraphics{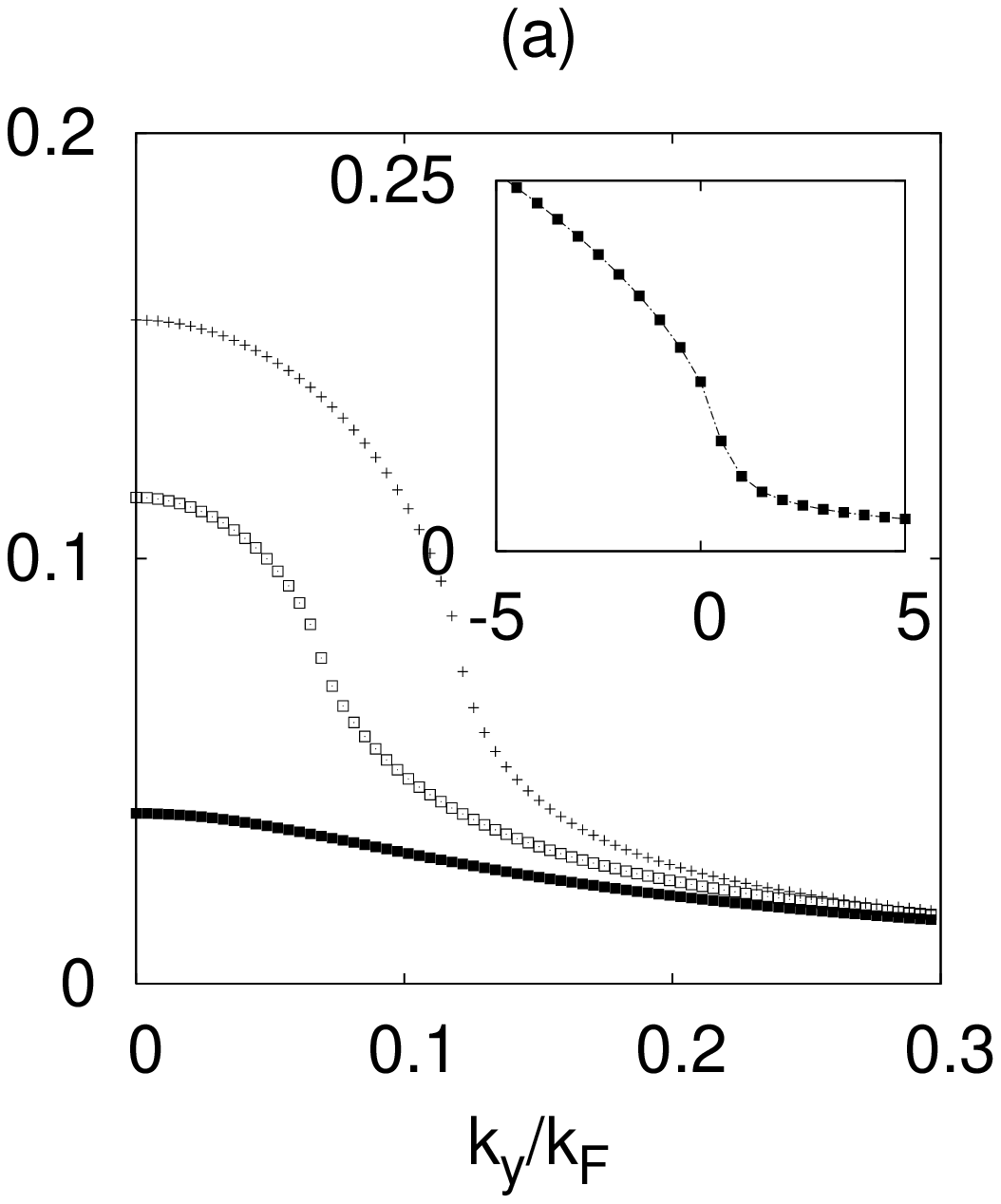} \includegraphics{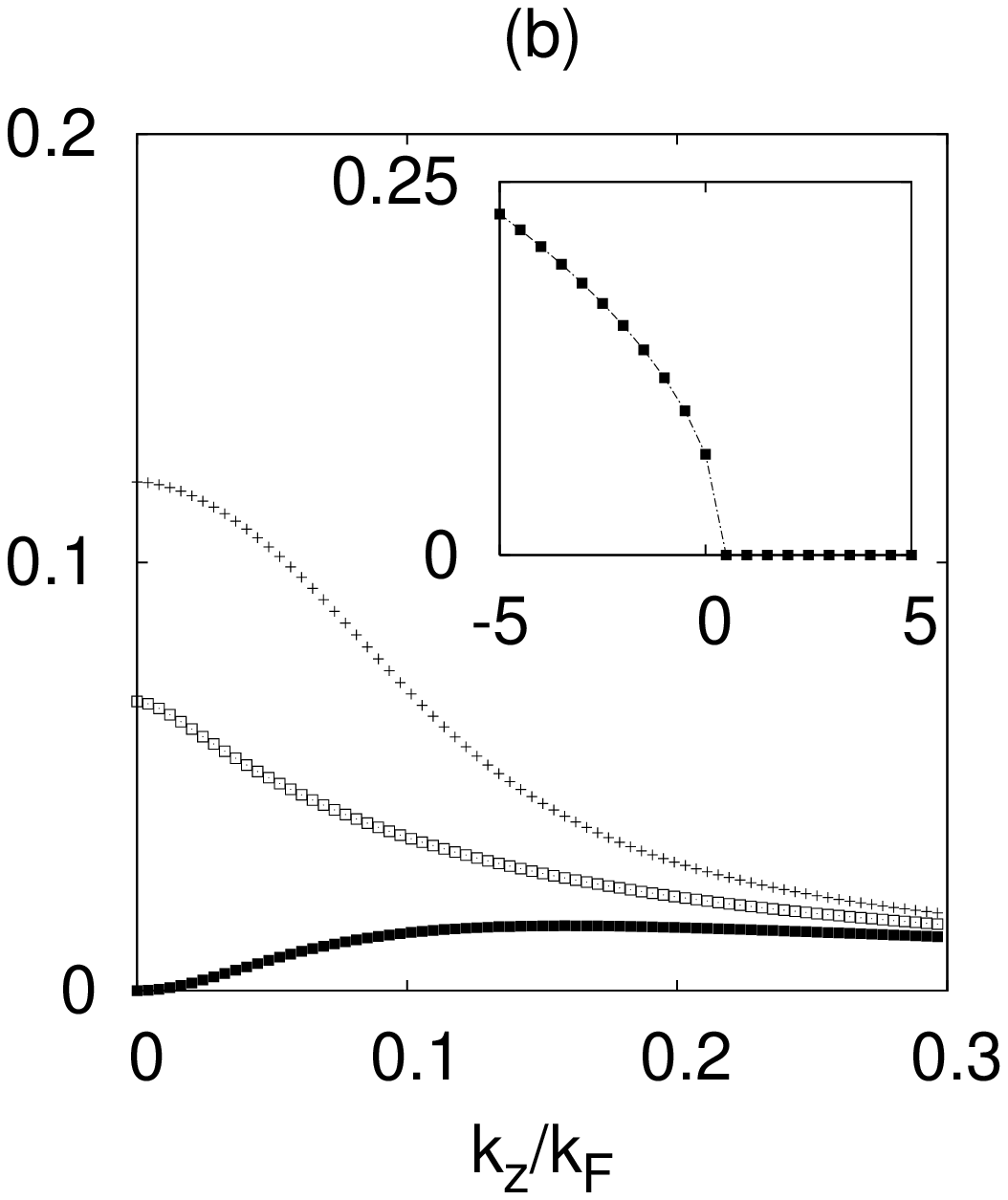}}}
\caption{\label{fig:md} Plots of integrated momentum distribution
(in units of $L_i k_{\rm F} /\pi$)
a) $n_{\rm z} (k_{\rm x} = 0, k_{\rm y})$ versus $k_{\rm y}/k_{\rm F}$
b) $n_{\rm y} (k_{\rm x} = 0, k_{\rm z})$ versus $k_{\rm z}/k_{\rm F}$
for $1/(k_{\rm F}^3a_p) = -1$ (+), $0$ (hollow squares) and $1$ (solid squares).
Insets show:
a) $n_{\rm z} (k_{\rm x} = 0, k_{\rm y} = 0)$;
b) $n_{\rm y} (k_{\rm x} = 0, k_{\rm z} = 0)$ 
versus $1/(k_{\rm F}^3 a_{\rm p})$.
}
\end{figure}

In Fig.~\ref{fig:md}, we show the integrated momentum distribution
$
n_{\rm i} (\mathbf{k}) = \sum_{\mathbf{k}_i} n_0(\mathbf{k}) = (L_i/2 \pi) \int dk_i n_0(\mathbf{k})
$
during the evolution from BCS to BEC, where $L_i$ is the length along ${i}^{\rm th}$ direction.
With increasing interaction strength, fermion pairs become more 
tightly bound, and thus, $n_i(\mathbf{k})$ becomes broader as fermions with 
larger momentum participate in the formation of bound states.
Notice also that, while the nodes of the order parameter are averaged over
upon $k_{\rm z}$ integration, they play an important role when either 
$k_{\rm y}$ or $k_{\rm x}$ is integrated.
To be specific, at $\mathbf{k} = 0$, $n_{\rm z}(0)$ decreases 
continuously as a function of coupling from BCS to BEC regime, 
and vanishes for $1/(k_{\rm F}^3 a_{\rm p}) \to \infty$.
However, $n_{\rm y}(0) = L_y (2M \mu)^{1/2}/\pi$ 
vanishes with coupling for $\mu > 0$ in the BCS side, 
and remains zero $n_{\rm y}(0) = 0$ for $\mu < 0$ in the BEC side.
Thus, the qualitative difference between $n_{\rm z}(\mathbf{k})$ and $n_{\rm y}(\mathbf{k})$ 
[or $n_{\rm x}(\mathbf{k})$] around $\mathbf{k} = 0$ explicitly 
shows a direct measurable consequence of the gapless to gapped 
quantum phase transition when $\mu = 0$.

Next we discuss, $p$-wave superfluidity near $T_{\rm c}$.
For $T = T_{\rm c}$ ($\Delta_0 = 0$),
$
N_{0} = \sum_{\mathbf{k}} n_{\rm F}[\xi(\mathbf{k})],
$
corresponds to the number of unbound fermions. 
Here, $n_{\rm F}(w) = 1/[\exp(\beta w) + 1]$ is the Fermi distribution.
The fluctuation contribution $N_{\rm fluct}$ is obtained as follows.
The matrix $\mathbf{F}^{-1}(q)$ can be simplified to yield 
\begin{equation}
L^{-1}(q) = \frac{1}{4\pi g} - \sum_{\mathbf{k}} 
\frac{1 - n_{\rm F}(\xi_+) - n_{\rm F}(\xi_-)} {\xi_+ + \xi_- - iv_\ell} |\Gamma(\mathbf{k})|^2,
\end{equation}
which is the generalization of the $s$-wave case~\cite{carlos}.
Here, $L^{-1}(q) = \mathbf{F}^{-1}_{11}(q)$, and $\xi_\pm = \xi(\mathbf{k} \pm \mathbf{q}/2)$.
The resulting action then leads to the thermodynamic potential
$\Omega_{\rm gauss} = \Omega_0 + \Omega_{\rm fluct}$, where
$
\Omega_{\rm fluct} = - \beta^{-1}\sum_{q} \ln[\beta L(q)].
$

The branch cut (scattering) contribution $\Omega_{\rm sc}$ to $\Omega_{\rm fluct}$ 
is obtained by writing $\beta L(q)$ in terms of the phase shift
$
\delta(\mathbf{q},w) = {\rm Arg} [\beta L(\mathbf{q},w + i0^+)],
$
leading to
$
\Omega_{{\rm sc}} = - \pi^{-1} \sum_{\mathbf{q}} \int_{w_q^*}^{\infty} n_{\rm B}(w) \widetilde{\delta}(\mathbf{q},w) dw,
$
where $w_\mathbf{q}^* = |\mathbf{q}|^2/(4M) - 2 \mu$ and
$\widetilde{\delta}(\mathbf{q},w) = \delta(\mathbf{q},w) - \delta(\mathbf{q},0)$.
Here, $n_{\rm B}(w) = 1/[\exp(\beta w) - 1]$ is the Bose distribution.
For each $\mathbf{q}$, the integral only contributes 
for $w > w_\mathbf{q}^*$, since $\delta(\mathbf{q},w) = 0$ otherwise.
Thus, the branch cut contribution to the number equation
$N_{{\rm sc}} = -\partial\Omega_{{\rm sc}} / \partial\mu$ is given by
\begin{eqnarray}
N_{{\rm sc}} = \frac{1}{\pi}\sum_{\mathbf{q}} \int_{0}^{\infty}
\left[ \frac{\partial n_{\rm B}(\widetilde w)}{\partial \mu} +  n_{\rm B}(\widetilde w) \frac{\partial}{\partial \mu} 
\right] \widetilde{\delta}(\mathbf{q},\widetilde w) dw,
\label{eqn:sc}
\end{eqnarray}
where $\widetilde w = w + w^*_{\mathbf q}$.

When $a_{\rm p} < 0$, there are no bound states above $T_{\rm c}$ and $N_{{\rm sc}}$ 
represents the correction due to scattering states.
On the other hand, when $a_{\rm p} > 0$, there may also be bound states in the two-fermion spectrum, 
represented by poles with $w < w_\mathbf{q}^*$.
For arbitrary $1/(k_{\rm F}^3 a_{\rm p})$,
the evaluation of the pole (bound state) contribution $N_{\rm bs}$ 
requires heavy numerics. However in strong coupling,
\begin{equation}
N_{{\rm bs}} = 2\sum_{\mathbf{q}} n_{\rm B}[w_\mathbf{q} - \mu_{\rm B}],
\label{eqn:bs}
\end{equation}
where $w_\mathbf{q} = |\mathbf{q}|^2/(4M)$ and $\mu_{\rm B} = -E_{\rm b} + 2\mu$.
Here, we use
$
1/(4\pi g) = \sum_{\mathbf{k}} |\Gamma(\mathbf{k})|^2 / [2\epsilon(\mathbf{k}) - E_{\rm b}]
$
to express Eq.~(\ref{eqn:bs}) in terms of binding energy $E_{\rm b} < 0$.
Notice that the expression for $N_{\rm bs}$ given above is good only for couplings where $\mu_{\rm B} < 0$.
Thus, our results for $k_0 \approx 200 k_{\rm F}$ are not strictly valid when 
$0 < 1/(k_{\rm F}^3 a_{\rm p}) < 1/(k_{\rm F}^3 a_{\rm p}^*) \sim 5$,
where $a_{\rm p}^*$ corresponds to $\mu_{\rm B} = 0$.
Therefore, in this region we interpolate.
The binding energy in the BEC regime is  $E_{\rm b} = -2/(M k_0 a_{\rm p})$ (when $k_0^3 a_{\rm p} \gg 1$).
This result is consistent with a $T$-matrix calculation~\cite{tlho}, 
where $E_{\rm b} = 1/(M a_{\rm p} r_{\rm p})$ with 
$
r_{\rm p} = - 2/(k_0^2 a_{\rm p}) - \pi k_0^2/(4 M^2 V) 
\sum_{\mathbf{k}} |\Gamma(\mathbf{k})|^2 / \epsilon^2(\mathbf{k}).
$
This leads to $r_{\rm p} = - k_0/2$ (when $k_0^3 a_{\rm p} \gg 1$), 
indicating that both approaches produce the same result. 

\begin{figure} [htb]
\centerline{\scalebox{0.37}{\includegraphics{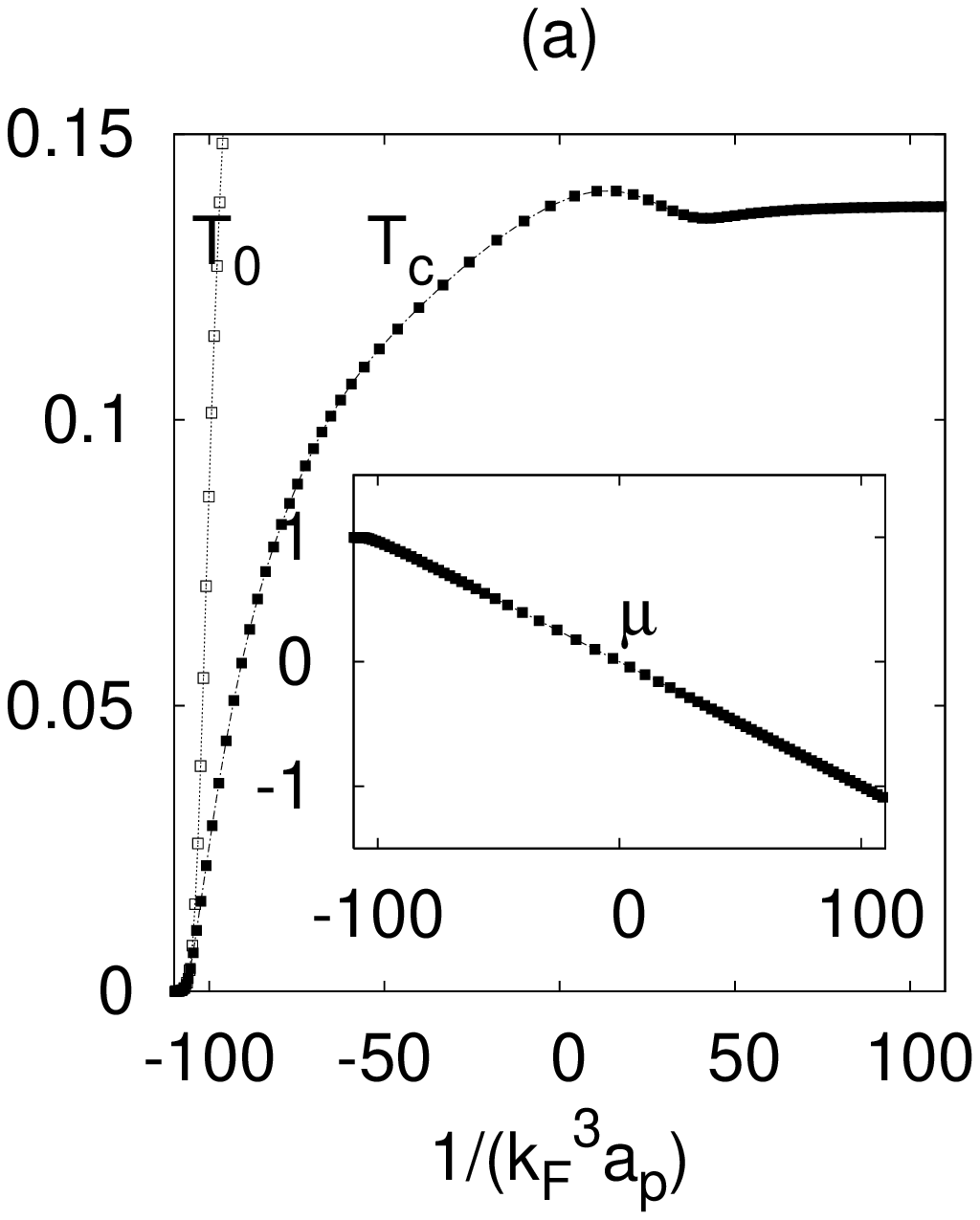} \includegraphics{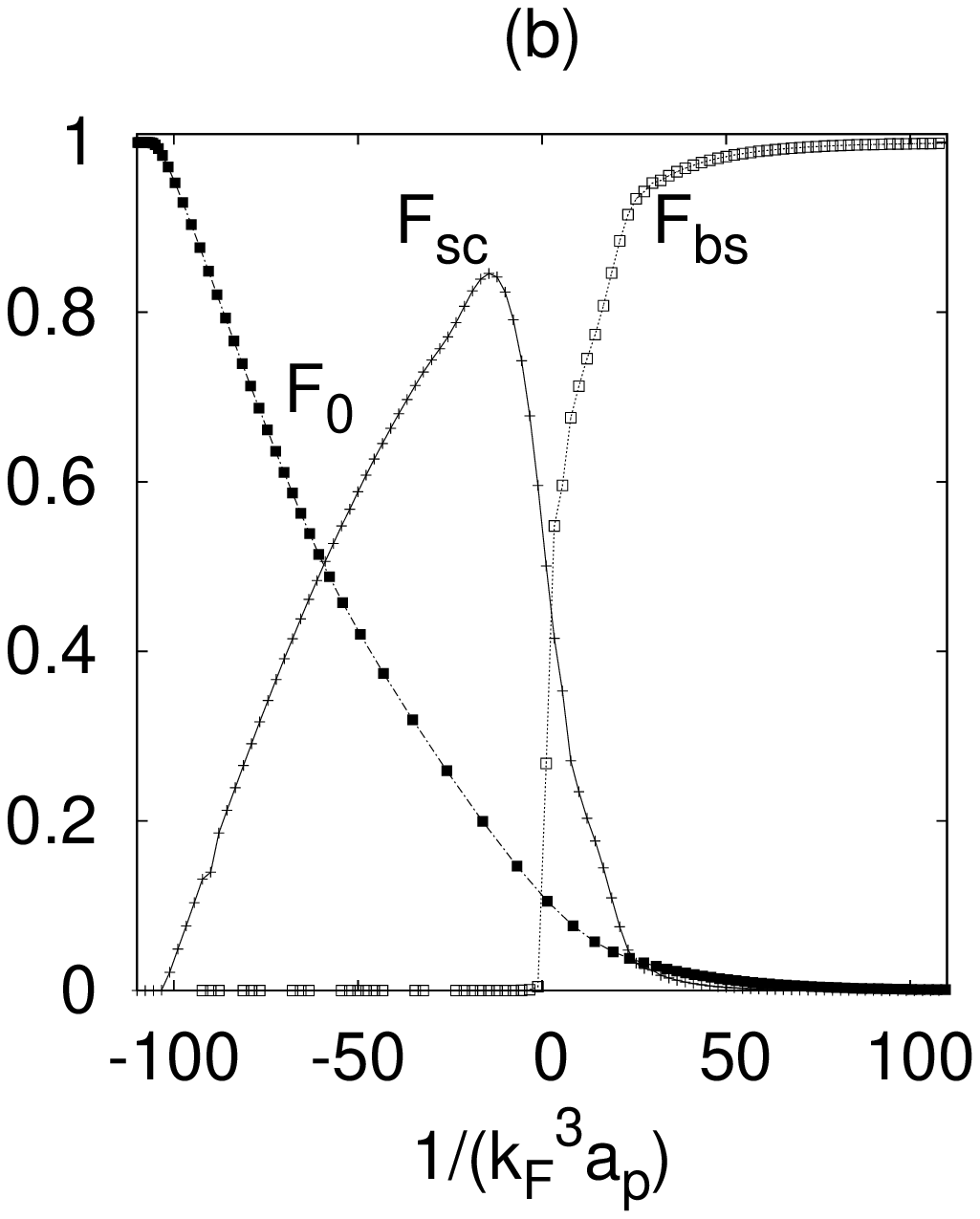}}}
\caption{\label{fig:tc.and.mu} Plots of reduced 
a) critical temperature $T_{\rm r} = T_{\rm c}/\epsilon_{\rm F}$ and
chemical potential $\mu_{\rm r} = \mu/\epsilon_{\rm F}$ (inset), and 
b) fraction of unbound $F_0 = N_0/N$, scattering $F_{\rm sc} = N_{\rm sc}/ N$, 
bound $F_{\rm bs} = N_{\rm bs} / N$ fermions at $T = T_{\rm c}$
versus $1/(k_{\rm F}^3 a_{\rm p})$. 
}
\end{figure}

To obtain the evolution from BCS to BEC, we solve numerically the
number $N = N_0 + N_{{\rm sc}} + N_{{\rm bs}}$ and order parameter equations.
In Fig.~\ref{fig:tc.and.mu}a, we plot $T_{\rm r} = T_{\rm c}/\epsilon_{\rm F}$ and $\mu_{\rm r} = \mu/\epsilon_{\rm F}$ 
as a function of $1/(k_{\rm F}^3 a_{\rm p})$.
The weak coupling 
$T_{{\rm c}} = (8/\pi)\epsilon_{\rm F} \exp [\gamma - 8/3 + \pi k_0 / (4k_{\rm F}) - \pi/(2k_{\rm F}^3|a_{\rm p}|)]$
evolves continuously to the dilute Bose gas
$T_{{\rm c}} = 2\pi [2 n_{\rm B}/\zeta(3/2)]^{2/3}/M_{\rm B} = 0.137\epsilon_{\rm F}$
in the BEC regime, where $\gamma \approx 0.577$ is the Euler's constant and 
$n_{\rm B} = n/2 = k_{\rm F}^3/(12 \pi^2)$ is the density and $M_B = 2M$ is the mass of the bosons.
However, the saddle point $T_{0} \approx E_{\rm b}/[2\ln(E_{\rm b}/\epsilon_{\rm F})^{3/2}]$ increases with 
$1/(k_{\rm F}^3 a_{\rm p})$, and is a measure of the pair dissociation temperature~\cite{carlos}. 
Notice that, the ratio of $\widetilde{\Delta} (k_{\rm F}) / T_{\rm c} = \Delta_0 \Gamma_{\rm k} (k_{\rm F})/T_{\rm c}$
in the BCS limit is $3\pi/e^\gamma$.
The hump in the intermediate regime is similar to the one observed in fermion-boson model~\cite{ohashi}.
Furthermore, similar humps were also calculated in the $s$-wave case~\cite{carlos},
however, whether they are physical or not may require a fully self-consistent numerical approach.

The weak coupling $\mu = \epsilon_{\rm F}$ evolves continuously to the 
strong coupling $\mu = -1/(M k_0 a_{\rm p})$ (when $k_0^3 a_{\rm p} \gg 1 $)
leading to $\mu = E_{\rm b}/2$. 
Notice that $\mu$ crosses the bottom of the band at $1/(k_{\rm F}^3 a_{\rm p}) \approx 0.5$, i.e., 
after the two-body bound state threshold $1/(k_{\rm F}^3 a_{\rm p}) = 0$ is reached. 
The evolution of
$\mu$ at $T = 0$ (Fig.~\ref{fig:gap.and.mu}) and $T = T_{\rm c}$ (Fig.~\ref{fig:tc.and.mu}) 
is similar, but very different from $s$-wave~\cite{carlos}.
However, another result for $\mu$ versus $1/(k_{\rm F}^3 a_{\rm p})$ at $T = T_{\rm c}$ 
(much like the $s$-wave case) was obtained in Ref.~\cite{ohashi} using a fermion-boson model.
In Fig.~\ref{fig:tc.and.mu}b, we also plot the fractions 
of unbound ($F_0 = N_0/ N$), scattering ($F_{\rm sc} = N_{{\rm sc}}/ N$),
and bound ($F_{\rm bs} = N_{{\rm bs}}/ N$) fermions as a function of $1/(k_{\rm F}^3 a_{\rm p})$.
While $N_0$ ($N_{{\rm bs}}$) dominates in weak (strong)
coupling, $N_{{\rm sc}}$ is dominant at the intermediate regime.

Next, we study the evolution of time-dependent Ginzburg-Landau (TDGL) equation near $T_{\rm c}$. 
We expand the effective action $S_{{\rm eff}}$ 
around $\Delta_0 = 0$ to fourth order in $\Lambda(q)$~{\cite{carlos}},
and obtain the TDGL equation
\begin{equation}
\left[ a + b|\Lambda(x)|^2 - \sum_{<i,j>}\frac{c_{ij}}{2M}\nabla_i\nabla_j - 
id\frac{\partial}{\partial t} \right]\Lambda(x) = 0,
\end{equation}
in the real space $x = (\mathbf{x},t)$ representation.
The time-independent expansion coefficients are given by
$
a = 1/(4\pi g) - \sum_{\mathbf{k}} X |\Gamma(\mathbf{k})|^2 / [2\xi(\mathbf{k})],
$
and
$
c_{ij} = \sum_{\mathbf{k}} \big\lbrace X\delta_{ij}/[8\xi^2(\mathbf{k})] - \beta Y\delta_{ij}/[16\xi(\mathbf{k})] 
+ \beta^2 XY k_i k_j /[16M\xi(\mathbf{k})] \big\rbrace 
|\Gamma(\mathbf{k})|^2,
$
where $\delta_{ij}$ is the Kronecker delta, 
$X = \tanh[\beta\xi(\mathbf{k})/2]$ and $Y = {\rm sech}^2[\beta\xi(\mathbf{k})/2]$.
The coefficient of the nonlinear term is
$
b = \sum_{\mathbf{k}} \big\lbrace X/[4\xi^3(\mathbf{k})] - \beta Y/[8\xi^2(\mathbf{k})] \big\rbrace 
|\Gamma(\mathbf{k})|^4.
$
The time-dependent coefficient has real and imaginary parts, and is given by
$
d = \sum_{\mathbf{k}} X|\Gamma(\mathbf{k})|^2/[4 \xi^2(\mathbf{k})]
+ i\beta N(\epsilon_{\rm F})\mu^{3/2}\Theta(\mu)/(32 \epsilon_0 \epsilon_{\rm F}^{1/2}),
$
where $\Theta (\mu)$ is the Heaviside function.
As the coupling grows, the coefficient of the propagating term 
(Re[$d$]) increases, while the damping term (Im[$d$]) decreases until it vanishes 
beyond $\mu = 0$. This indicates that the dynamics of $\Lambda(x)$ is undamped
for $\mu < 0$. For completeness, we present next the asymptotic forms of $a, b, c_{ij}$ and $d$.

In weak coupling ($\mu = \epsilon_{\rm F}$), we find
$a = \kappa_{\rm w} \ln(T/T_{\rm c})$,
$b = 2\kappa_{\rm w} \epsilon_{\rm F}\zeta(3) / (5\pi T_{\rm c}^2\epsilon_0)$,
$c_{xx} = c_{yy} = c_{zz}/3 = 7\kappa_{\rm w} \epsilon_{\rm F} \zeta(3) / (20\pi^2 T_{\rm c}^2)$, 
$c_{i \ne j} = 0$, and
$d = \kappa_{\rm w} [1/(4\epsilon_{\rm F}) + i\pi/(8T_{\rm c})]$,
where $\kappa_{\rm w} = \epsilon_{\rm F} N(\epsilon_{\rm F}) / (4\pi\epsilon_0)$ and 
$\zeta(x)$ is the zeta function.
By rescaling the order parameter
$\Psi_{\rm w}(x) = \sqrt{b/\kappa_{\rm w}}\Lambda(x)$
one obtains the conventional TDGL equation 
$
\epsilon \Psi_{\rm w} + |\Psi_{\rm w}|^2\Psi_{\rm w} - \sum_i (\xi_{{\rm GL}}^{ii})^2 \nabla_i^2\Psi_{\rm w} + \tau_{{\rm GL}}\partial_t\Psi_{\rm w} = 0
$
with characteristic length $\xi_{ii}^2 = c_{ii}/(2M a) = (\xi_{{\rm GL}}^{ii})^2 / \epsilon$ 
and time $\tau = -id/a = \tau_{{\rm GL}}/\epsilon$ scale. 
Here, $\epsilon = (T_{\rm c} - T)/T_{\rm c}$ with $|\epsilon| \ll 1$, 
$k_{\rm F}\xi_{{\rm GL}}^{xx} = k_{\rm F}\xi_{{\rm GL}}^{yy} = 
k_{\rm F}\xi_{{\rm GL}}^{zz}/3 = \sqrt{7\zeta(3)/(20\pi^2)} (\epsilon_{\rm F}/T_{\rm c})$,
and $\tau_{{\rm GL}} = -i/(4\epsilon_{\rm F}) + \pi/(8T_{\rm c})$. 
The system is overdamped since $T_{\rm c} \ll \epsilon_{\rm F}$ reflecting the 
presence of two-fermion continuum states into which Cooper pairs can decay.

In strong coupling ($\epsilon_0 \gg |\mu| \gg T_{\rm c}$), we find
$a = \kappa_{\rm s}(2|\mu| - |E_{\rm b}|)/8$,
$b = 9\kappa_{\rm s} / (256\pi\epsilon_0)$,
$c_{ij} = \kappa_{\rm s} \delta_{ij} / 16$, and
$d = \kappa_{\rm s} / 8$,
where $\kappa_{\rm s} = N(\epsilon_{\rm F}) / (4\sqrt{\epsilon_{\rm F}\epsilon_0})$.
By rescaling the order parameter
$\Psi_s(x) = \sqrt{d}\Lambda(x)$
one obtains the conventional Gross-Pitaevskii equation for a dilute gas of bosons
$
\mu_{\rm B} \Psi_{\rm s} + U_{\rm B}|\Psi_{\rm s}|^2\Psi_{\rm s} - \nabla^2\Psi_{\rm s}/(2M_{\rm B}) - i\partial_t\Psi_{\rm s} = 0
$
with bosonic chemical potential $\mu_{\rm B} = - a/d = 2\mu - E_{\rm b}$, mass $M_{\rm B} = M d/c_{ii} = 2M$, 
and repulsive interactions $U_{\rm B} = b/d^2 = 18\pi/(M k_0)$. 
In this regime, $k_{\rm F} \xi_{{\rm GL}}^{ii} = [\pi k_0/(36k_{\rm F})]^{1/2}$ is independent
of $a_{\rm p}$, and is infinitely large when $k_0/k_{\rm F} \to \infty$.

The evolution of $\xi_{{\rm GL}}^{ii}$ follows from
$
(\xi_{{\rm GL}}^{ii})^2 = c_{ii} / [2M T_{\rm c}(\partial a / \partial T)]
$
where
$
\partial a / \partial T = \sum_{\mathbf{k}} \big\lbrace Y/(4T^2) + 
(\partial \mu / \partial T) [Y/(4T\xi(\mathbf{k})) - X/(2\xi^2(\mathbf{k})) ] \big\rbrace |\Gamma(\mathbf{k})|^2.
$
Notice that, $\partial \mu/\partial T$ vanishes in weak coupling, 
while it plays an important role in strong coupling.
The evaluation of $\partial \mu / \partial T$ for intermediate coupling is very difficult, thus
an interpolation for $\xi_{{\rm GL}}^{zz}$ connecting the weak and strong coupling regimes 
is shown in Fig.~\ref{fig:gap.and.mu}b.
While $\xi_{\rm GL}^{ii}$ representing the phase coherence length is large compared to interparticle 
spacing in both BCS and BEC limits, it has 
a minimum in the unitarity region $1/(k_{\rm F}^3 a_{\rm p}) \approx 0$.

In the same figure, we also compare $\xi_{{\rm GL}}^{zz}$ and the average Cooper pair size 
$
\xi_{{\rm pair}}^2 = - \langle \psi(\mathbf{k})| \nabla_\mathbf{k}^2 |\psi(\mathbf{k}) \rangle / 
\langle \psi(\mathbf{k}) | \psi(\mathbf{k}) \rangle,
$
where $\psi(\mathbf{k}) = \Delta(\mathbf{k})/[2E(\mathbf{k})]$ 
is the $T = 0$ pair wave function.
Notice that $\xi_{{\rm pair}}$ is a decreasing function of interaction (while $\xi_{{\rm GL}}^{ii}$
is not). The limiting value of $\xi_{\rm pair}$ in strong coupling is controlled by $k_{\rm F}/k_0$. 
Furthermore, $\xi_{{\rm pair}}$ has a cusp (non-analiticity) when $\mu = 0$. 
This cusp is associated with the change in $E(\mathbf{k})$ from gapless 
(with line nodes) in the BCS to fully gapped in the BEC side.  

In conclusion, we analysed the evolution of superfluid properties of a 3D
dilute $p$-wave Fermi gas from weak (BCS) to strong (BEC) coupling regime as a 
function of scattering volume at temperatures $T = 0$ and $T = T_{\rm c}$. 
We discussed the order parameter, chemical potential, average Cooper pair size, 
and momentum distribution at the ground state ($T = 0$).
We also discussed the critical temperature $T_{\rm c}$, chemical potential and number of 
unbound, scattering and bound fermions at the normal state ($T = T_{\rm c}$).
Lastly, we derived the TDGL equation for $T \approx T_{\rm c}$
and extracted the GL coherence length.

We thank NSF (DMR-0304380) for support.


\begin{thebibliography}{99}

\bibitem{regal} C. A. Regal et al., Phys. Rev. Lett. \textbf{90}, 053201 (2003).
\bibitem{ticknor} C. Ticknor et al., Phys. Rev. A \textbf{69}, 042712 (2004).
\bibitem{zhang} J. Zhang et al., Phys. Rev. A \textbf{70}, 030702 (2004).
\bibitem{schunck} C. H. Schunck et al., Phys. Rev. A \textbf{71}, 045601 (2005).
\bibitem{gunter} K. G\"unter et al., cond-mat/0507632.
\bibitem{john} J. L. Bohn, Phys. Rev. A \textbf{61}, 053409 (2000).
\bibitem{gurarie} V. Gurarie et al., Phys. Rev. Lett. \textbf{94}, 230403 (2005).
\bibitem{skyip} C. -H. Cheng and S. -K. Yip, Phys. Rev. Lett. \textbf{95}, 070404 (2005).
\bibitem{tlho} T. -L. Ho and R. B. Diener,  Phys. Rev. Lett. \textbf{94}, 090402 (2005).
\bibitem{botelho} S. S. Botelho and C.A.R. S{\'a} de Melo, J.L.T.P. \textbf{140}, 409 (2005).
\bibitem{iskin} M. Iskin and C.A.R. S{\'a} de Melo, cond-mat/0508134.
\bibitem{ohashi} Y. Ohashi, Phys. Rev. Lett. \textbf{94}, 050403 (2005).
\bibitem{carlos} C. A. R. S{\'a} de Melo et al., Phys. Rev. Lett. \textbf{71}, 3202 (1993).

\end{thebibliography}
\end{document}